\documentclass[conference]{IEEEtran}
\IEEEoverridecommandlockouts
\usepackage{cite}
\usepackage{amsmath,amssymb,amsfonts}
\usepackage{algorithmic}
\usepackage{graphicx}
\usepackage{textcomp}
\usepackage{xcolor}
\usepackage{multirow}
\usepackage{todonotes}
\usepackage[symbol]{footmisc}

\def\BibTeX{{\rm B\kern-.05em{\sc i\kern-.025em b}\kern-.08em
    T\kern-.1667em\lower.7ex\hbox{E}\kern-.125emX}}
\begin{document}

\title{Fault Detection Effectiveness of Metamorphic Relations Developed for Testing  Supervised Classifiers}

\author{\IEEEauthorblockN{Prashanta Saha}
\IEEEauthorblockA{\textit{School of Computing, Montana State University}\\
Bozeman, USA \\
prashantasaha@montana.edu}
\and
\IEEEauthorblockN{Upulee Kanewala* \thanks{*Corresponding author}}

\IEEEauthorblockA{\textit{School of Computing, Montana State University} \\
Bozeman, USA\\
upulee.kanewala@montana.edu}
}

\maketitle

\begin{abstract}
In machine learning, supervised classifiers are used to obtain predictions for unlabeled data by inferring prediction functions using labeled data. Supervised classifiers are widely applied in domains such as computational biology, computational physics and healthcare to make critical decisions. However, it is often hard to test supervised classifiers since the expected answers are unknown. This is commonly known as the \emph{oracle problem} and metamorphic testing (MT) has been used to test such programs. In MT, metamorphic relations (MRs) are developed from intrinsic characteristics of the software under test (SUT). These MRs are used to generate test data and to verify the correctness of the test results without the presence of a test oracle. Effectiveness of MT heavily depends on the MRs used for testing. In this paper we have conducted an extensive empirical study to evaluate the fault detection effectiveness of MRs that have been used in multiple previous studies to test supervised classifiers. Our study uses a total of 709 reachable mutants generated by multiple mutation engines and uses data sets with varying characteristics to test the SUT. Our results reveal that only 14.8\% of these mutants are detected using the MRs and that the fault detection effectiveness of these MRs do not scale with the increased number of mutants when compared to what was reported in previous studies.

\end{abstract}

\begin{IEEEkeywords}
Metamorphic testing, Random testing, Supervised classifiers, Metamorphic Relations, Mutation Analysis, Machine Learning
\end{IEEEkeywords}

\section{Introduction}
\label{sec:intro}
Supervised classifiers are widely used for making predictions in diverse domains. For instance, over fifty real world computational applications use support vector machines for classification \cite{svm}. As these types of applications are becoming part of our daily life, ensuring their quality becomes even more important \cite{Michalski:2013:MLA:2588013}.
In such applications, formal proofs of the underlying algorithm does not always guarantee that it implements that algorithm correctly. Therefore, software testing is imperative to assure the quality of these systems.

Often, conventional software testing approaches may not be feasible for assuring the quality of supervised classifiers because of the absence of a test oracle that determines the correctness of produced test outputs. This class of software applications is often referred to as ``non-testable programs'' \cite{doi:10.1093/comjnl/25.4.465}. Further, usually supervised classifiers are not 100\% accurate. Thus, an incorrect prediction does not necessarily mean that there is a fault in the program. These characteristics of supervised classifiers make it hard to detect subtle faults in these applications. 

To date, limited work has been done on systematic testing of software systems that incorporate machine learning. Among them, metamorphic testing (MT) has been used widely for testing software applications that uses supervised machine learning algorithms \cite{Murphy2008PropertiesOM,Xie:2011:TVM:1942318.1942371,Dwarakanath:2018:IIB:3213846.3213858,Ding:2017:VDL:3103620.3103631}. 
MT uses metamorphic relations (MRs) for testing the software under test (SUT), where MRs act as partial oracle \cite{Chen2003FaultbasedTW,Chen:2018:MTR:3177787.3143561}. A MR specifies how the outputs should change according to a specific change made to the input. Thus, from existing test cases (named as \emph{source test cases}) MRs are used to generate new test cases (named as \emph{follow-up test cases}).
If the changes found between the outputs of the source and follow-up test cases are not as expected according to the MR, then there is a defect in the SUT. Thus, using MRs we can address the oracle problem presented by supervised machine learning classifiers.

Several previous studies have defined MRs for testing supervised classifiers. Xie et al. proposed a set of MRs based on user expectations to validate supervised classifiers \cite{Xie:2011:TVM:1942318.1942371}. Dwarakanath et al. developed MRs for two image classifiers that are based on support vector machines and  deep learning \cite{Dwarakanath:2018:IIB:3213846.3213858}. Ding et al. developed three levels of MRs to test and validate a deep learning framework \cite{Ding:2017:VDL:3103620.3103631}. The evaluations conducted in these studies to measure the fault detection effectiveness of the developed MRs is fairly limited due to the number of mutants used. For example Xie et al. used 24 mutants and Dwarakanath et al. used 22 mutants in total.
These numbers are significantly low especially considering the number of classes and the number of lines of source code involved with these SUTs.


To overcome this limitation, in this paper, we report the findings of a large scale experiment that we conducted to evaluate the fault detection effectiveness of MRs developed for supervised classifiers. In this experiment we used a total of 709 reachable mutants (i.e. the mutated statement in the mutant was executed with the test cases) that were generated for a real world supervised classifier implementation from Weka \cite{witten05:_data_minin}. Our results show a significant reduction of the fault detection effectiveness with the increased number of mutants. 

Rest of the paper organized as follows: Section~\ref{sec:back} describes the background of this work, including an overview of supervised machine learning and the k-Nearest Neighbors algorithm, which is used as the SUT in this study. Section~\ref{sec:mt} discusses more about MT and the MRs used for testing. In Section~\ref{sec:exp} we discuss the details of our experimental approach and mutation analysis. Section~\ref{sec:result} presents the results and their analysis. Section~\ref{sec:rel} identifies the related work and Section~\ref{sec:con} contains our conclusions and future work.
\section{Background}
\label{sec:back}
\subsection{Supervised Machine Learning Classifiers}
\label{sec:super}
Supervised classification is the task of deducing a function from labeled training data such that it can be used to predict unknown labels on test data. Training data can be represented by two vectors of size $k$. One vector is the training samples $S = <s_0,s_1,...,s_{k-1}>$ and the other one is the class labels $C =<c_0,c_1,..,c_{k-1}>$ where, $c_i$ is the class label for $s_i$. Each sample $s_i$ has $m$ features from which the prediction function will be learned. Class labels are a finite set and each class label $c_i$ is an element of it,  i.e. $c\in L = <l_0,l_1,...,l_{n-1}>,$ where $n$ is the number of class labels \cite{Xie:2011:TVM:1942318.1942371}.

Supervised machine learning applications execute in two phases: the \emph{training phase} and \emph{testing phase}. In the \emph{training phase}, a set of training samples are used by a supervised classification algorithm to learn a prediction function. To develop the prediction function, the supervised learning algorithm would analyze how the attributes relate to the class label. In the \emph{testing phase}, the prediction function is applied to unseen data known as the \emph{test set}, where the class labels are unknown. The application attempts to predict the class label for each instance in the test set using the learned prediction function \cite{Xie:2011:TVM:1942318.1942371}. Some of the commonly used supervised classification algorithms are K-Nearest Neighbors \cite{Cover:2006:NNP:2263261.2267456}, Naive Bayes \cite{Rish01anempirical} and Support Vector Machine \cite{svm}.

\subsection{K-Nearest Neighbors}
\label{sec:knn}
 This study uses an open-source implementation of the K-Nearest Neighbors (kNN) algorithm as the SUT. kNN is particularly chosen due to its popularity in the machine learning community and is used in domains such as recommendation systems, semantic searching and anomaly detection etc. Further, Xie et al. used kNN in their study and using the same algorithm would allow us to do a comparison with their results \cite{Xie:2011:TVM:1942318.1942371}. However, the MT approach discussed here should be applicable to other supervised learning algorithm implementations.


In kNN, for a sample training set $S$, each sample set has $m$ attributes,$<att_0,att_1,...,att_{m-1}>$, and also $n$ classes,$<l_o,l_1,...,l_{n-1}>$. The sample test data is $t_s$,$<a_0,a_1,...,a_{m-1}>$.
kNN computes the distance between each sample training set and the test case. Euclidean distance metric is one of the most popular approach to measure distance. For sample $s_i \in S$, the value for each attribute is $<sa_0,sa_1,...,sa_{m-1}>$. And the euclidean distance formula is:\\

$\displaystyle dist(s_i,t_s)=\sqrt{\sum_{j}^{m-1}\left(sa_j-a_j\right)^2}$

Once the distance is calculated, kNN selects the $k$ nearest training samples for the test data after sorting all the distances. These $k$ samples from the training set are considered as the \emph{k-nearest neighbors} of the test case. Then, kNN calculates the proportion of each label in the selected k-nearest neighbors. The class label with the highest proportion is predicted as the label for the test data.
\section{Metamorphic Testing for supervised classifiers}
\label{sec:mt}
Often, programs exhibit properties such that if the test input is changed in a way that the new output can be predicted based on the original output. In MT, such properties (known as MRs) are used as partial oracles to conduct testing \cite{Chen:2015:MTS:2819261.2819278, 6319263}. In practice one can easily apply MT. As the first step, it is necessary to identify MRs that can relate multiple pairs of the inputs and outputs of the SUT.
Then, source test cases are generated using techniques like random testing, structural testing or search based testing and the corresponding follow-up test cases are constructed based on the MRs. In our previous studies we investigated how the fault detection effectiveness of MT varies with various source test case generation techniques such as different structural coverage based approaches and our results show that coverage based source test case generation outperforms randomly generated source test cases~\cite{Saha:2018:FDE:3193977.3193982}. After executing the source and the follow-up test cases on the SUT we can check if there is a change in the output that matches the MR, if not the MR is considered as violated. Violation of a MR during testing indicates faults in the SUT. Since MT checks relationship between inputs and outputs of a test program, we can use this technique when the expected result of individual test output is unknown.

 %
For example, \textit{Consistence with affine transformation} MR described in Section~\ref{sec:MRs} can be used to test a kNN classifier. Source test case for kNN can be randomly generated (see Table~\ref{tab:sampledata} for an example - training data on the left and test data on the right). After executing this source test case, the output will be the class label predicted for that test case which is 0 for this example. To generate follow-up test case, we apply the input transformation described in the above MR, where an arbitrary affine transformation function is applied to the attributes of both the training data and test data. After executing the follow-up test case the output is 0 which is the predicted class label for the transformed test data. To satisfy this MR both the source and follow-up test case outputs should be same. Therefore, in this example, the considered MR is satisfied for the given source and follow-up test cases.

\subsection{Identified MRs for Testing kNN}
\label{sec:MRs}
Murphy et al. \cite{Murphy2008PropertiesOM} suggested six MRs (additive, multiplicative, permutative, invertive, inclusive and exclusive) that can be applied to machine learning applications including both supervised and unsupervised ML. Xie et al. developed a set of MRs based on the user expectations of supervised classifiers \cite{Xie:2011:TVM:1942318.1942371}. In our study, we mainly use the MRs developed by Xie et al. to test kNN. Some variations of the same MRs are used in some recent studies as well~\cite{Dwarakanath:2018:IIB:3213846.3213858}.  
Below we provide a brief description of these MRs (formal definitions can be found in~\cite{Xie:2011:TVM:1942318.1942371}).

\textbf{MR1: Consistence with affine transformation.}
 If we apply some affine transformation function, $f(x) = kx + b, (k \neq 0)$, to every value $x$ in some subset of features in the training and testing data, and then create a new model using this data, the predictions made by the model should be unchanged.

\textbf{MR2: Permutation of the attribute.}
If we permute the order of the attributes, or features, of all the samples in the training and testing data, the result of the predictions of the test data should not change.

\textbf{MR3: Addition of uninformative attributes.}
If we add some new feature that is equally associated with all classes, the predictions of the test data should not be changed.

\textbf{MR4: Consistence with re-prediction.}
Suppose we predict some test case $t$ as class $l_i$. If we append $t$ to our training data and re-create the model, $t$ should still be classified as class $l_i$.

\textbf{MR5: Additional training sample.}
Suppose we predict some test case $t$ as class $l_i$. If we duplicate all samples of class $l_i$ in our training data and re-classify our test data, $t$ should still be classified as $l_i$. More generally, every test case predicted as class $l_i$ should still be predicted as class $l_i$ with the duplicated samples.

\textbf{MR6: Addition of classes by re-labeling samples.}
For some test cases not of class $l_i$, we switch the class label from $x$ to $x^{*}$. Then every test case predicted as class $l_i$ should still be predicted as class $l_i$ with the re-labeled samples.

\textbf{MR7: Permutation of class labels.}
 If we permute the order of the class labels with some random permutation $p(l_i)$ where $l_i$ is a class label, all test cases which were predicted as $l_i$ should now be predicted as $p(l_i)$.

\textbf{MR8: Addition of informative attribute.}
If we add some new feature that is strongly associated with one class, $l_i$, then for every prediction that was class $l_i$, the prediction with this new attribute should also be class $l_i$.

\textbf{MR9: Addition of classes by duplicating samples.}
Suppose we duplicate every class except for $n$, and give them all a new class. For example, if we originally had class labels of 1, 2, 3, and 4, then we would create class labels of 1, 1*, 2, 2*, 3, 4, 4*. Then every test case predicted as class $l_i$ (class 3 in this example) should still be predicted as class $l_i$ with the duplicated samples.

\textbf{MR10: Removal of classes.}
If we remove some class $l_i$, the remaining predictions should remain unchanged.

\textbf{MR11: Removal of samples.}
If we remove samples that have not been predicted as class $l_i$, then all cases which were predicted as $l_i$ should remain unchanged.

When using these MRs for testing kNN, it is important to select the appropriate value for $k$ (i.e. number of nearest neighbours) such that the MR becomes a necessary property for kNN. Table \ref{MRs} shows the $k$ values used with each MR for verification of kNN \cite{Xie:2011:TVM:1942318.1942371}.
\begin{table}[ht]
\caption{Sample Data Set}
\label{tab:sampledata}
\begin{tabular}{lllll}
\cline{1-2}
\multicolumn{1}{|l|}{\begin{tabular}[c]{@{}l@{}}@attribute pictures numeric\\ @attribute paragraphs numeric\\ @attribute files numeric\\ @attribute files2 numeric\\ @attribute profit \{0,1,2,3,4\}\\ \\ @data\\ 45,3,16,38,0\\ 15,87,89,46,4\\ 59,77,94,11,0\\ 86,89,94,15,2\\ 80,28,94,11,4\\ 23,12,47,41,1\\ 94,15,22,15,0\\ 95,26,97,76,3\\ 50,90,0,72,2\\ 33,46,47,95,0\end{tabular}} & \multicolumn{1}{l|}{\begin{tabular}[c]{@{}l@{}}@attribute pictures numeric\\ @attribute paragraphs numeric\\ @attribute files numeric\\ @attribute files2 numeric\\ @attribute profit \{0,1,2,3,4\}\\ \\ @data\\ 6,40,8,89,0\end{tabular}} &  &  &  \\ \cline{1-2}

\end{tabular}
\end{table}
\section{Experimental Study}
\label{sec:exp}
The goal of this experimental study is to conduct an in-depth evaluation of the fault detection effectiveness of the MRs listed in Section~\ref{sec:MRs}.  We used the kNN implementation in Weka 3.5.7  as the SUT \cite{witten05:_data_minin}. 
\subsection{Research Questions}
\label{sec:rq}
We conducted a set of experiments to answer the following research questions:
\begin{enumerate}
\item \textbf{ How does the fault detection rate of MRs vary as the number of mutants increase?} As we mentioned above the fault detection effectiveness of these MRs were measured using a limited set of mutants in previous studies \cite{Xie:2011:TVM:1942318.1942371}. But it is important to use a reasonable number of mutants to evaluate the fault detection effectiveness using the mutation killing rate. Therefore we increase the number of mutants significantly and measure the killing rate.
\item \textbf{Does the fault detection rate of MRs change with varying  the  data  set  size  in  the source  test  cases?} There are two major components in MT that determines its fault detection effectiveness: the MRs and the source test cases. We examined whether varying the data set size of the source test case can effect the mutant killing rate for a given MR.
\item \textbf{Does the fault detection effectiveness vary with the mutation engine used to generate the mutants?} As we discussed above the underlying process used by MuJava and Major for generating mutants is different. The purpose of this research question is to see whether that one category of mutants are hard to kill than the other.
\end{enumerate}
\subsection{Source and Follow-up Test Cases}
\label{sec:testcase}
In an individual source test case
there is a training set and a test set. Similar to Xie et al.\cite{Xie:2011:TVM:1942318.1942371}
, we used a random approach to generate these source test cases. In each training and test set, there are four numerical attributes that are named as: $\{pictures, paragraphs, files, files2\}$. The class label $profit$ can have five values $\{0,1,2,3,4\}$.
The value of each attribute is randomly selected and ranges within [0,~100]. The value of the class labels are also selected randomly. The training set size ranges within [10,~200]. Table \ref{tab:sampledata} shows a sample source test case, with the training data set on the left and test data set on the right.


We transform the source test cases to obtain the corresponding follow-up test cases according to the MRs described in Section~\ref{sec:MRs}. 
Multiple source and follow-up test case pairs are generated for each MR by varying the number of samples in the training data set as well as the size of source test case. 

We executed all the source and follow-up test case pairs on the original kNN implementation and validated the outputs against each MR before proceeding to the mutation analysis described below. The original kNN implementation did not report any MR violations.
\begin{table}[ht]
\caption{Metamorphic relations for kNN used in mutation analysis}
\label{MRs}
\begin{tabular}{|l|p{0.4\textwidth}|}
\hline
k=3 & \begin{tabular}[c]{@{}l@{}}MR1 Consistence with affine transformation\\ MR2 Permutation of the attribute\\ MR3 Addition of uninformative attributes\\ MR4 Consistence with re-prediction\\ MR5 Additional training sample\\ MR6 Addition of classes by re-labeling samples\end{tabular} \\ \hline
k=1 & \begin{tabular}[c]{@{}l@{}}MR7 Permutation of class labels\\ MR8 Addition of informative attribute\\ MR9 Addition of classes by duplicating samples\\ MR10 Removal of classes\\  MR11 Removal of samples\end{tabular}                          \\ \hline
\end{tabular}
\end{table}
\subsection{Mutation Analysis}
\label{sec:mut}
To evaluate the fault detection effectiveness of the MRs described in Section~\ref{sec:MRs}, we use a mutation engine to systematically inject defects into the SUT. Mutation testing has been extensively used to evaluate fault detection effectiveness, as many experiments suggest that mutants are a proxy to the real faults for comparing testing techniques \cite{1553583}. As we mentioned above, previous studies used  mutation analysis to evaluate the fault detection effectiveness of MRs \cite{Xie:2011:TVM:1942318.1942371}. But, the number of mutants used in the mutation analysis is quite low compared to the size of the SUT's used in those experiments.

In our evaluation, we applied MuJava \cite{Ma:2005:MAC:1077303.1077304} and Major \cite{Just:2014:MMF:2610384.2628053} tools to systematically generate mutants for kNN in Weka-3.5.7. MuJava is a powerful and automatic mutation analysis system, which can support both method-level and class-level mutation operators. MuJava provides various types of mutants, including inter-class, intra-class, inter-method and intra-method level of mutants. In this experiment, we only included the intra-method level of mutants.
Major is a mutation testing framework which manipulates the abstract syntax tree of the SUT. Similar to MuJava, we only used the intra-method level mutant operators from the Major tool. 

MuJava and Major allows users to define which parts of the source code needs to be mutated. Since, Weka is a large scale software with about 16.4M source code and our experiments only focuses on the functionality of the kNN classifier, we only selected the class files which are directly related to the kNN classifier according to its hierarchy structure. Table \ref{tab:files} shows the names of the selected class files in our mutation analysis.
\begin{table}[ht]
\caption{Selected files for mutation analysis}
\label{tab:files}
\begin{tabular}{|p{0.45\textwidth}|}
\hline
kNN \\ \hline
\begin{tabular}[c]{@{}l@{}}weka.classifiers.lazy.IBk.java\\ weka.core.Attribute.java\\ weka.core.neighboursearch.LinearNNSearch.java\\weka.core.neighboursearch.NearestNeighbourSearch.java\\weka.core.NormalizableDistance.java\\weka.core.EuclideanDistance.java \end{tabular} \\ \hline
\end{tabular}
\end{table}
We generated all possible mutants for the 6 class files in Table~\ref{tab:files}. After excluding the mutants that caused compilation errors, runtime exception as well as equivalent mutants, we have obtained a total of 1500 mutants from the MuJava and Major mutation tools. From those mutants we identified 609 mutants generated by MuJava that are reachable by the test cases that we described above. From the mutants generated by Major, we randomly selected 100 mutants that are reachable by the test cases due to time limitations. The distribution of mutants between the two tools are described in Table \ref{mutants}.



\begin{table}[ht]
\caption{Details of Mutants }
\label{mutants}
\begin{tabular}{|p{0.1\textwidth}|p{0.15\textwidth}|p{0.15\textwidth}|}
\hline
Tool Name & \begin{tabular}[c]{@{}c@{}}Total \# of  mutants \\ generated\end{tabular} & \# of mutants used \\ \hline
MuJava & 2383 & 609 \\ \hline
Major & 987 & 100 \\ \hline
\end{tabular}
\end{table}

\section{Results and Discussions}
\label{sec:result}

Below we discuss the results of our experiments and provide answers to our research questions.
\begin{figure*}[ht]
  \includegraphics[scale=0.8]{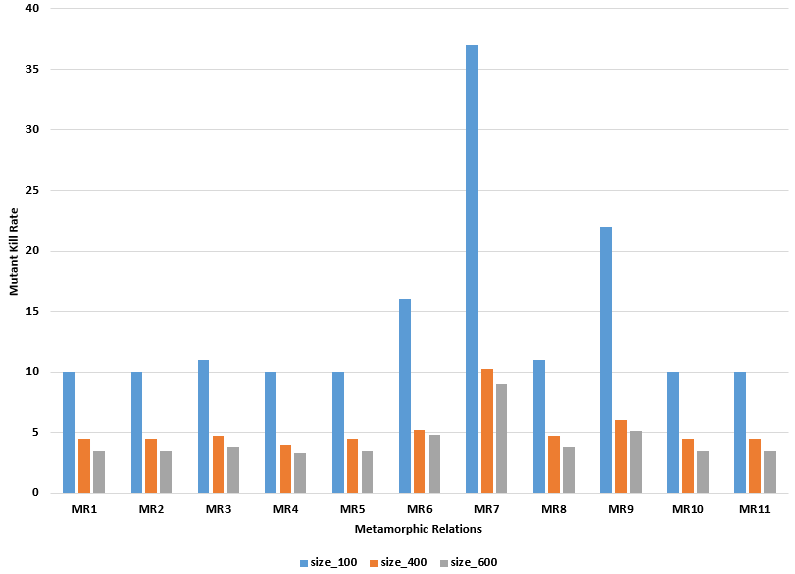}
  \caption{Mutant kill rate for each MR by varying Mutant Size.}
  \label{fig2_2}
\end{figure*}

\textbf{1. How does the fault detection rate of MRs vary as the number of mutants increase?}

Out of the 709 mutants (609 MuJava + 100 Major) used in this experiment, only 105 
(14.8\%) mutants could be killed using the MRs. This is a significant decrease in the mutation killing rate compared to what is  reported in Xie et al., where 19 out of the 24 (79\%) mutants were killed by the same set of MRs~\cite{Xie:2011:TVM:1942318.1942371}. We think that the reason for this significant decrease in the mutation killing rate is due to the fact that Xie et al. used a selected set of mutation operators to generate the mutants used in their experiment and those mutants do not provide a good representation of the potential faults in the SUT.

To further evaluate how the mutation killing rate varies when increasing the number of mutants, we executed the MRs with mutant sets of size 100, 400 and 600 that are randomly selected from the mutants generated by the MuJava tool. We used 10 randomly generated source test cases for executing each of the MRs. Figure \ref{fig2_2} shows the mutant kill rates for each MR for the three mutant sets. 
As, shown in Figure \ref{fig2_2}, mutant kill rate for all the MRs reduced when the number of mutants were increased. In particular, the killing rates for sizes 400 and 600 are significantly lower compared to size 100 for MR6, MR7, and MR9.\\

\fbox{\begin{minipage}{23em}
 A Significant decrease in the mutation killing rate with the increased number of mutants.
\end{minipage}}
\\ 

\textbf{2. Does the fault detection rate of MRs change with varying  the  data  set  size  in  the source  test  cases?}

 The goal of this research question is to identify whether fault detection effectiveness of MRs vary with the size of the data sets used as the source test case. In this experiment, we created data sets of 18 different sizes where the number of samples vary from 30 to 200. These data sets were executed on 100 mutants that were randomly selected from the MuJava mutants. 
 
\begin{figure*}[ht]
  \includegraphics[scale=0.85]{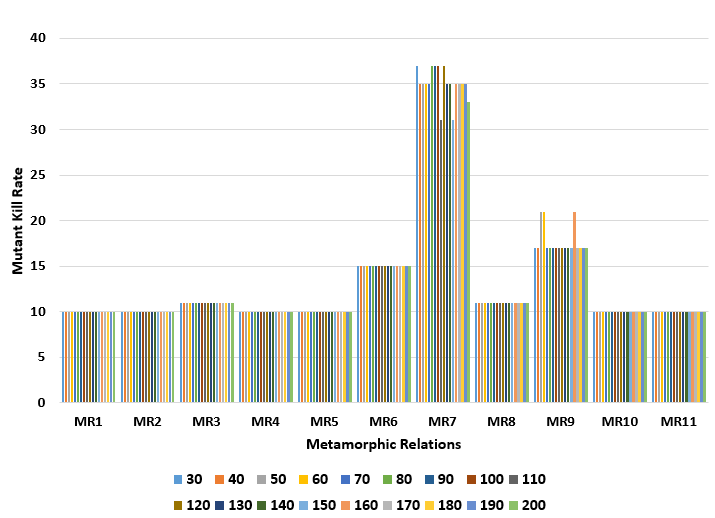}
  \caption{Mutant kill rate by each MR in kNN with varying data Set.}
  \label{fig2}
\end{figure*}

 Figure \ref{fig2} shows the mutant killing rate for each MR with varying number of samples in the source test cases. It is interesting to note that mutants killing rate is low for all the MRs across the different data sets ranging between 10\% and 37\%. Only MR7 and MR9 is showing some variation in the killing rate which is 6\% and 4\%, respectively. On the other hand, the rest of MRs have a constant mutants killing rate despite the difference in data set sizes used as the source test cases. In summary, varying the size of the random sample data has no significant effect on the fault detection effectiveness of the considered MRs. But it might be possible to increase the fault detection effectiveness by generating test data based on some test coverage criterion as we discussed in our previous study \cite{Saha:2018:FDE:3193977.3193982}.\\
 
 \fbox{\begin{minipage}{23em}
 No major changes in the kill rate with varying data set size.
\end{minipage}}
\\

\textbf{3. Does the fault detection effectiveness vary with the mutation engine used to generate the mutants?}

\begin{figure*}[ht]
  \includegraphics[scale=0.85]{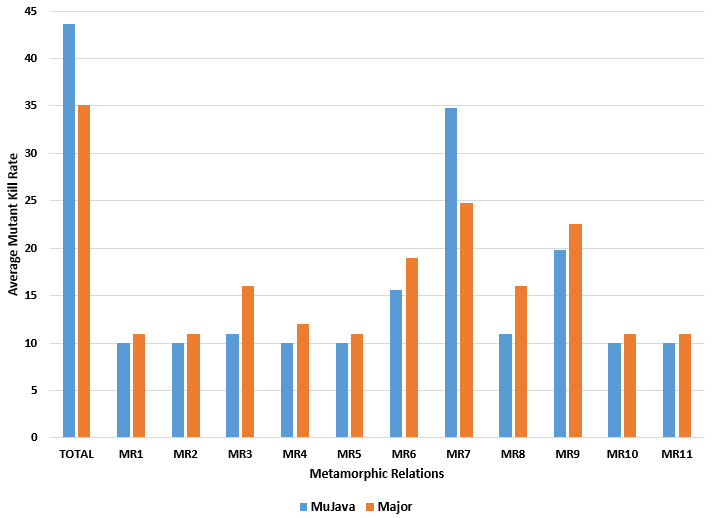}
  \caption{Average mutants kill rate by each MR for MuJava and Major tool.}
  \label{fig3}
\end{figure*}

In order to answer this research question, we used mutants from MuJava and Major mutation tools. We used 10 randomly generated sample data sets as source test cases for each MR and executed them on a set of 100 randomly selected mutants from MuJava and a set of 100 randomly selected mutants from Major. We report the results of this evaluation in Figure \ref{fig3}.

As shown in Figure \ref{fig3}, the overall mutant killing rate on the MuJava and Major mutants is 43.6\% and 35.1\%, respectively. When comparing the results at the individual MR level, it is noticeable that there is some consistency in the killing rate for each MR between these two tools. For example, for both tools, MR7 and MR9 have comparatively higher mutants killing rate than the other MRs. Also it is interesting to note that for all the MRs except MR7, killing rate of Major mutants is higher than that of the MuJava mutants even though the overall killing rate is higher for MuJava mutants. This indicates that the mutation killing is dominated by MR7. In summary, overall MuJava mutants are easier to kill, while with majority of MRs Major mutants are easier to kill.\\

\fbox{\begin{minipage}{23em}
 Most effective MRs perform consistently across the two mutation tools.
\end{minipage}}
\\
\section{Related Work}
\label{sec:rel}
There has been a significant amount of work done on applying machine learning to solve software testing problems compared to testing machine learning applications \cite{4601522}. For example, machine learning has been used to predict likely MRs for a given programs\cite{kanewala2015Pred,KanewalaBiemanISSTA13,Rahman:2018:PMR:3193977.3193983,Hardin:2018:USL:3193977.3193985}.

MT has been applied to test different types of machine learning applications \cite{Murphy2008PropertiesOM}. A case study done on real world machine learning application framework shows that MRs can effectively detect faults \cite{Xie:2011:TVM:1942318.1942371}. A recent work \cite{Dwarakanath:2018:IIB:3213846.3213858}  investigated the application of metamorphic testing to test complex machine learning algorithms such as SVMs with non-linear kernels and deep residual neural networks (ResNET). The technique was able to successfully detect mutants in open-source machine learning applications.

MRs has been proven to be a core element of MT. In image processing applications MT was used by Tahir et al. \cite{7176238}. They have shown that only few MRs that are related to specific images are more effective in detecting faults than others. Regardless of conducting MT, MRs have been  used for the augmentation of the machine learning models \cite{8457613}. Here MRs were identified based on properties of the input data and the usage of the binary classification model. 
Hui et al. \cite{7552281} has proposed a semi automated MT approach for GIS testing that used the superficial area calculation program to illustrate the process of the testing approach. They have developed a MR model to generate compound MRs. 

Some research efforts are reported on how to identify effective MRs. Asrafi et al. \cite{5992013} have observed a correlation between the test code coverage achieved by an MR and its fault detection effectiveness. In object oriented software testing a method of constructing MRs based on algebraic specification has been proposed \cite{7515931}. This method provides low MRs redundancy and improves the efficiency of software testing. $\mu$MT \cite{7811324} a MR construction tool that uses data mutation to construct an input relation and the generic mapping rule associated with each mutation operator to construct output relation.
\section{Conclusions and Future work}
\label{sec:con}
Previous studies have developed MRs for conducting MT on supervised classifiers. But, a major drawback of these studies is the limited number of mutants used to evaluate their fault detection effectiveness. In this paper, we empirically evaluated the fault detection effectiveness of MRs developed for supervised classifiers using a set of 709 reachable mutants, which is a significant increase in the number of mutants compared to what is used in the previous studies. \


Our study shows that the MRs identified based on user expectations of supervised classifiers are not as effective in detecting faults as claimed in previous studies. Out of the 709 mutants only 14.8\% of mutants could be detected using these MRs. Our study also shows that changing the size of randomly generated data used as source test cases does not have an effect on the fault detection effectiveness of these MRs. 

In the future, we plan to develop MRs based on specific algorithmic properties of commonly used supervised classifiers. We think such MRs will have higher fault detection effectiveness compared to the ones we investigated in this study. We also plan to investigate ways to develop more effective source test cases for this domain using various data distributions. Further, we plan to extend this experiment to other machine learning algorithms including deep learning algorithms.

\section*{Acknowledgment}
This work is supported by award number 1656877 from the National Science Foundation. Any Opinions, findings and conclusions or recommendations expressed in this material are those of the author(s) and do not necessarily reflect those of the National Science Foundation.

\bibliographystyle{IEEEtran}

\end{document}